\documentclass[aps,pra,amsmath,amssymb,amsfonts,twocolumn,superscriptaddress,a4paper]{revtex4-2}

\usepackage{graphicx}
\usepackage{braket}
\usepackage[colorlinks=true,linkcolor=blue,citecolor=blue,urlcolor=blue]{hyperref}
\usepackage{cleveref} 

\date{\today}

\begin{document}
	
\title{Network nonlocality sharing via weak measurements in the generalized star network configuration}

\author{Jian-Hui Wang}
\author{Ya-Jie Wang}
\author{Liu-Jun Wang}
\email{ljwangq@ynu.edu.cn}
\author{Qing Chen}
\email{chenqing@ynu.edu.cn}
\affiliation{School of Physics and Astronomy and Yunnan Key Laboratory for
	Quantum Information, Yunnan University, Kunming 650500, China}

\begin{abstract}
	Network nonlocality exhibits completely novel quantum correlations compared to standard quantum nonlocality. It has been shown that network nonlocality can be shared in a generalized bilocal scenario via weak measurements [\href{https://journals.aps.org/pra/abstract/10.1103/PhysRevA.105.042436https://journals.aps.org/pra/abstract/10.1103/PhysRevA.105.042436}{Phys. Rev. A. {\bf 105}, 042436 (2022)}].  In this paper, we investigate network nonlocality sharing via weak measurements in a generalized star-shaped network configuration with arbitrary numbers of unbiased dichotomic input $k$, which includes $n$ branches and adds ($m$-1) more parties in each branch to the original star network $(n, m=1, k=2)$ scenario. It is shown that network nonlocality sharing among all observers can be revealed from simultaneous violation of $2^n$ inequalities in the ($n, m=2, k=2$) and ($n, m=2, k=3$) scenarios for any $n$ branches. The noise resistance of network nonlocality sharing with a precise noise model is also analyzed.
\end{abstract}

\maketitle

\section{Introduction}
Quantum nonlocality represents a kind of quantum correlation that differs from the classical consideration. This concept was mathematically materialized by John Bell \cite{1964Bell} based on local hidden variable (LHV) models and it can be observed in experiments by violating some specific Bell inequalities. As a resource, quantum nonlocality has been used to implement secure key distribution \cite{2014BrunnerN,2007Antonio,2014Vazirani,2012Barrett} and randomness generation \cite{2010Pironio,2012Antonio,2017Curchod,2021Foletto}.

Different from standard quantum nonlocality, network nonlocality based on the independent sources assumption has been used to find many distinct phenomena, such as correlations that are compatible with the standard LHV models but incompatible with the network LHV models \cite{2010Branciard,2012Branciard,2017Gisin,2017France}. Furthermore, the exploration of network nonlocality has been extended to structures with different topologies \cite{2012FritzTo,2014Tavakoli,2015Mukherjee,2017Mukherjee,2019Marc,2020Mukherjee,2021Yang,2022Marc,2022Munshi,2022AlexP}, some of which have been  observed in experiments \cite{2017Gonzalo,2017DylanJ,2017Francesco,2019PanJ,2020FabioS,2022E,2022Huang} (see Ref. \cite{2022ArminTa} for a review).

Most works of the Bell scenario focus on one pair of entangled particles distributed to only two separated observers, Alice and Bob. If we divide one of the observers, say Alice, into a series of independent Alices that sequentially measure the same particle and all observers perform weak or unsharp measurements except the last Alice and Bob, who each perform strong measurements, then these observers may be able to simultaneously share nonlocality (nonlocality sharing). This concept was first proposed by Silva \cite{2015Silva}, and the works in nonlocality sharing have mainly concentrated on the sequential case on one side for recycling an arbitrarily long sequence of Alices \cite{2015Silva,2016MalS,2019DasD,2019KumariA,2020BrownP,2021ZhangT,2022Anna} or active nonlocality sharing \cite{2019RenCL,2020FengTF}. This kind of theory of sharing has been generalized to other quantum correlations, such as steering \cite{2018Sasmal,2019Akshata,2021Yao,2021Gupta,2022ZhuJie}, entanglement \cite{2018Bera,2020Foletto,2020Maity} and multipartite nonlocality \cite{2019Saha,2021ZhangT,2021Ren}. Recently, studies on nonlocality sharing have been extended to two-sided scenarios with two-qubit Clauser-Horne-Shimony-Holt (CHSH) inequality \cite{1969Clauser} in unbiased measurement selection but the two-sided nonlocality sharing was not found \cite{2021Cheng,2022Cheng,2022ZhuJie}.

 Hou \emph{et al.} first studied network nonlocality sharing in a two-sided situation based on the bilocal scenario, one of the simplest networks, and observed that four network inequalities, with respect to two Alices and two Bobs, can be simultaneously violated  \cite{2022Hou}. More recently, Mao \emph{et al.} experimentally observed network nonlocality sharing in a star-shaped network \cite{2022Mao}. These two works are both restricted to the two-setting scenario. An increasing number of inputs provides advantages in device-independent protocols \cite{2018Lee}. In this paper, we aim to explore network nonlocality sharing in an $n$-branch generalized star network scenario with $m$ observers in each branch and $k$ settings per observer, called the $(n,m,k)$ scenario. The chained $n$-locality inequality \cite{2018Lee,2020FabioS}, which requires only separable measurements performed by the center observer Bob, is used. Network nonlocality sharing among all observers can be revealed from simultaneous violation of $2^n$ inequalities in the $(n,2,2)$ and $(n,2,3)$ scenarios. The noise resistance of network nonlocality sharing with a precise noise model is also analyzed.

This paper is structured as follows: In Sec. \ref{S2}, we introduce the generalized $n$-branch star network model and the chained $n$-locality inequality for the star network. In Sec. \ref{S3}, the quantum upper bound for any $(n,m,k)$ star network scenario is derived. In Sec. \ref{S4}, star network sharing is discussed in detail under the optimal relation of weak measurement parameters. In Sec.\ref{S5}, noise resistance of star network sharing with a precise noise model is analyzed, and it is summarized in Sec.\ref{S6}.

\section{Generalized star network model and chained $\textit{\textbf{n}}$-locality inequality}\label{S2}

\begin{figure}[tpb]
	\centering
	\includegraphics[width=0.4\textwidth]{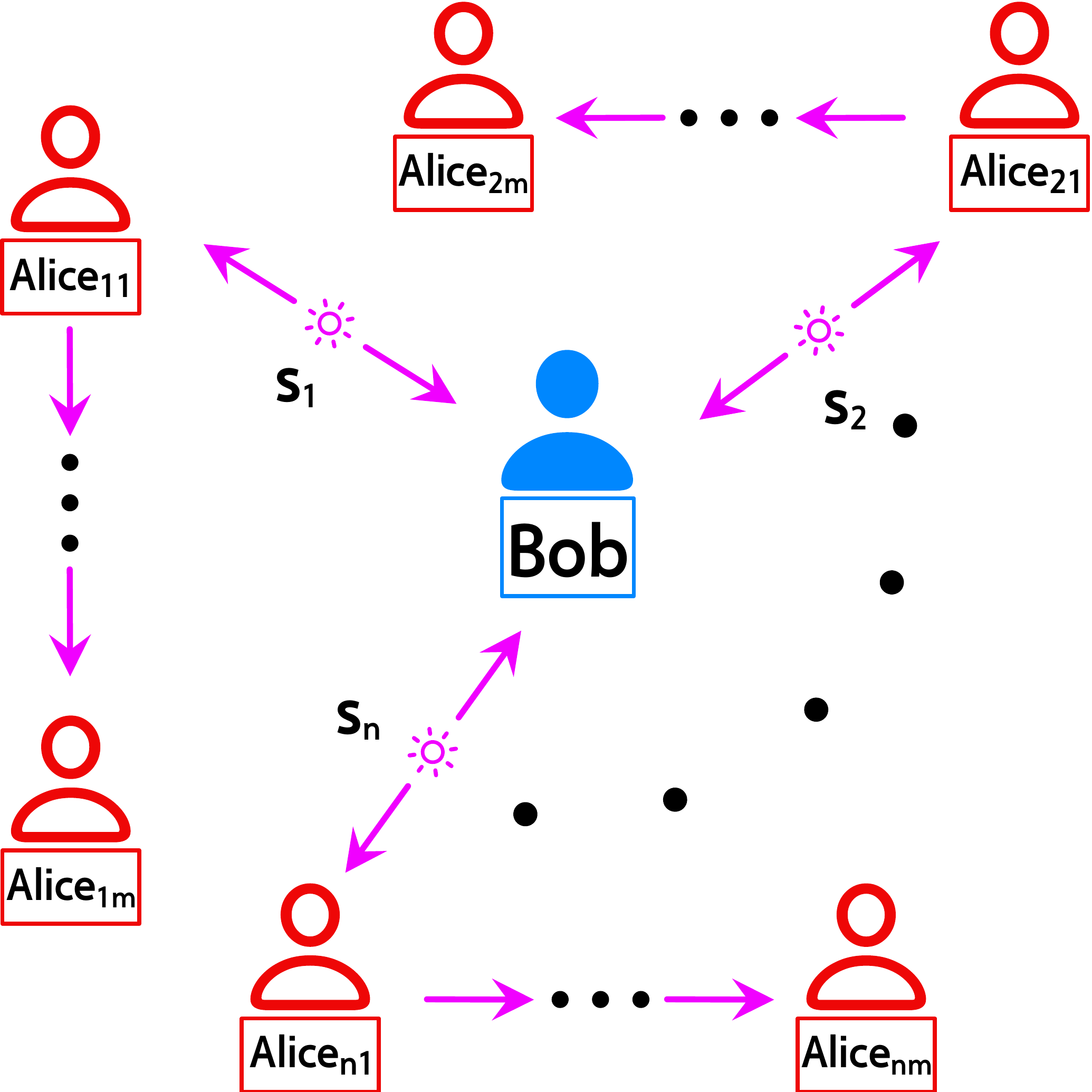}
	\caption{\small{The generalized star network configuration.  The original star network only includes all $\mathrm{Alice}_{i1}, i\in\{1,\dots,n\}$ and Bob. In each branch $i$, $\mathrm{Alice}_{i1}$ shares with Bob a two-qubit state produced by source $\mathrm{s}_i$. In the generalized star network configuration, which is different from the original star network, the first external observer $\mathrm{Alice}_{i1}$ performs weak measurements of the received particle and transmits it to the next Alice who also performs weak measurements, until transimited to the last $\mathrm{Alice}_{im}$ who performs strong measurements.}}\label{F1}
\end{figure}

The generalized star network nonlocality scenario, where $n$ independent sources $\mathrm{s}_1,\dots,\mathrm{s}_n$ distribute a two-qubit state to the corresponding $\mathrm{Alice}_{11},\dots,\mathrm{Alice}_{n1}$ and the center Bob, is described in Fig. \ref{F1}. The inputs of Bob and the peripheral $\mathrm{Alice}_{ij^{(i)}}$ are denoted by $y\in\{0,\dots,k-1\}$ and $x_{ij^{(i)}}\in\{0,\dots,k-1\}$, respectively, and their outcomes are denoted by $b=b_1\dots b_n\in\{1,-1\}$ and $a_{ij^{(i)}}\in\{1,-1\}$, respectively, where $i\in\{1,\dots,n\}$ and $j^{(i)}\in\{1,\dots,m\}$. The first subscript of $\mathrm{Alice}_{ij^{(i)}}$ denotes the $i_{\mathrm{th}}$ branch and the second subscript tells us the $j^{(i)}_{\mathrm{th}}$ observer in a given branch $i$.

The $n$-partite distribution
$P(a_{1j^{(1)}},\dots,a_{nj^{(n)}},b\mid x_{1j^{(1)}},\dots,x_{nj^{(n)}},y)$ is network $n$-local if it can be written in the  following factorized form

\begin{eqnarray}\label{E1}
	&&P(a_{1j^{(1)}},\dots,a_{nj^{(n)}},b\mid x_{1j^{(1)}},\dots,x_{nj^{(n)}},y)\nonumber\\
	&&=\int\left(\prod_{i=1}^{n}d\lambda_iq_{i}(\lambda_{i}) \ p(a_{ij^{(i)}}\mid x_{ij^{(i)}},\lambda_{i})\right) \\
	&&\times p(b | y, \lambda_{1}, \dots,\lambda_{n}),\nonumber
\end{eqnarray}

where $q_i(\lambda_i)$ is the distribution of the hidden variable $\lambda_i$ and the $n$ sets of distributions of hidden variable $\lambda_{1},\dots,\lambda_{n}$ originate from $n$ independent sources. For brevity, we write $\lambda=\lambda_1\cdot\cdot\cdot\lambda_{n}$. The local response function for $\mathrm{Alice}_{ij^{(i)}}$ only depends on $\lambda_{i}$ for any given $j^{(i)}$, and that of Bob depends on $\lambda$. 

The star network $n$-local model admits the following nonlinear chained $n$-locality Bell inequality \cite{2018Lee,2020FabioS}:

\begin{eqnarray}\label{E2}
	&&S^{(n,m,k)}_{j}=\sum\limits_{l=1}^k\vert \mathcal{I}_{lj}\vert^{1/n}\leq k-1,\\ 
	&&\mathrm{where}\ \mathcal{I}_{lj}=\nonumber\\
	&&\frac{1}{2^n}\sum\limits^l_{x_{1j^{(1)}},\dots,x_{nj^{(n)}}=l-1}\langle A_{1j^{(1)}}^{x_{1j^{(1)}}}\dots A_{nj^{(n)}}^{x_{nj^{(n)}}}B^{l-1}\rangle,\nonumber
\end{eqnarray}

where $A^k_{ij^{(i)}}=-A^0_{ij^{(i)}}$, $A^{x_{ij^{(i)}}}_{ij^{(i)}}$ ($B^l$) denotes the observable of $\mathrm{Alice}_{ij^{(i)}}$ (Bob) with input $x_{ij^{(i)}}$ ($l$) and $j=j^{(1)}\dots j^{(n)}$ denotes the sequentially-involved external observers; e.g., $S^{(3,4,5)}_{321}$ represents a (3,4,5) scenario corresponding to a three-branch star network with four Alices in each branch, and every observer in the network has five inputs and involves $\mathrm{Alice}_{13},\mathrm{Alice}_{22}$, $\mathrm{Alice}_{31}$ and a center Bob.

This inequality is a generalization of the original star network $n$-locality inequality \cite{2014Tavakoli}, the bilocal inequality \cite{2012Branciard} and the chained CHSH inequality \cite{1990Samuel}. When the maximally entangled state is shared in every branch, the quantum upper bound is $C_k=k\mathrm{cos}(\pi/2k)$ \cite{2018Lee}, which only depends on the number of inputs $k$.

When the singlet state $\ket{\psi^-}=(\ket{01}-\ket{10})/\sqrt{2}$ is shared in each branch, the measurements of the external observers are given as

\begin{equation}\label{E3}
	A^{x_{ij^{(i)}}}=\mathrm{sin}(\frac{x_{ij^{(i)}}\cdot\pi}{k})\sigma_{x}+\mathrm{cos}(\frac{x_{ij^{(i)}}\cdot\pi}{k})\sigma_{z},
\end{equation}

where $\sigma_{x}$ and $\sigma_{z}$ are Pauli matrices, and the positive-operator-value measure (POVM) elements of $\mathrm{Alice}_{ij^{(i)}}$ are given by $M_{a_{ij^{(i)}}\mid x_{ij^{(i)}}}=\frac{\mathbb{I}+a_{ij^{(i)}}A^{x_{ij^{(i)}}}}{2}$.

The center observer Bob performs a separable measurement with a product format $B^y=B^{y}_1\otimes\dots\otimes B^{y}_n$, where $B^{y}_i$ represents the measurement performed on the $i_{\mathrm{th}}$ subsystem:

\begin{equation}\label{E4}
B^{y}_i=\mathrm{sin}\frac{(2y+1)\pi}{2k}\sigma_{x}+\mathrm{cos}\frac{(2y+1)\pi}{2k}\sigma_{z}.
\end{equation}

The POVM elements of $\mathrm{Bob}$ are given by $M_{b\mid y}=M_{b_{1}\mid y_{1}}\otimes \dots\otimes M_{b_{n}\mid y_{n}}$ and $M_{b_{i}\mid y_{i}}=\frac{\mathbb{I}+b_{i}B^{y}_i}{2}$.

Recall that the operator corresponding to Bob's measurements factorizes as $B^y=B^{y}_1\otimes\dots\otimes B^{y}_n$, resulting in $\langle A_{1j^{(1)}}^{x_{1j^{(1)}}},\dots,A_{nj^{(n)}}^{x_{nj^{(n)}}}B^{y}\rangle=\prod\limits_{i=1}^n\langle A_{ij^{(i)}}^{x_{ij^{(i)}}}B^{y}_i\rangle$.

\section{Quantum upper bound of chained $\textit{\textbf{n}}$-locality inequalities for the $(\textit{\textbf{n,m,k}})$ scenario}\label{S3}

To calculate the quantum upper bound of Eq. \eqref{E2} for any $n,m,k$, we first need to calculate the correlator 

	\begin{eqnarray}\label{E5}
		&&\langle A_{ij^{(i)}}^{x_{ij^{(i)}}}B^{y}_i\rangle=\frac{1}{k^{(j^{(i)}-1)}}\\
		&&\times \sum\limits_{a_{ij},x_{ij},b_i}a_{ij^{(i)}}b_iP(a_{i1},\dots,a_{ij^{(i)}},b_i\mid x_{i1},\dots, x_{ij^{(i)}},y_i),\nonumber
	\end{eqnarray}

where $a_{ij}=a_{i1},\dots,a_{ij^{(i)}}$ and $x_{ij}=x_{i1},\dots,x_{i(j^{(i)}-1)}$.

To obtain the distribution $P(a_{i1},\dots,a_{ij^{(i)}},b_i\mid x_{i1},\dots, x_{ij^{(i)}},y_i)$, the following steps are required. 

Denote the state that $\mathrm{Alice}_{i1}$ shares with Bob as $\rho_{i}$.
Bob performs single-bit measurements with outcome $b^i$ on the particle he receives, this subsystem will change to

\begin{eqnarray}\label{E6}
\rho_{A_iB_i}^{b_i}=(\mathbb{I}\otimes M_{b_i\mid y_i})\cdot \rho_{i}\cdot(\mathbb{I}\otimes M_{b_i\mid y_i})^\dagger.
\end{eqnarray}

We do not normalize $\rho_{A_iB_i}^{b_i}$ because the probability distribution can be directly obtained by tracing the final unnormalized state. After Bob's measurement, the reduced state on $\mathrm{Alice}_{i1}$ can be obtained by tracing out Bob's system

\begin{eqnarray}\label{E7}
\rho^{b_i}_{A_i}=\mathrm{Tr}_B(\rho_{A_iB_i}^{b_i}).
\end{eqnarray}

If $j^{(i)}<m$, $\mathrm{Alice}_{ij^{(i)}}$ performs weak measurements on her subsystem. According to the discussion in Ref. \cite{2015Silva}, the reduced state can be given in the following recursive formula

\begin{eqnarray}\label{E8}
&&\rho_{ij^{(i)}}=\frac{F_{ij^{(i)}}}{2}\rho_{i(j^{(i)}-1)}\\
&&+\frac{1+a_{ij^{(i)}}G_{ij^{(i)}}-F_{ij^{(i)}}}{2}[M_{1\mid x_{ij^{(i)}}}\rho_{i(j^{(i)}-1)}(M_{1\mid x_{ij^{(i)}}})^\dagger]\nonumber \\
&&+\frac{1-a_{ij^{(i)}}G_{ij^{(i)}}-F_{ij^{(i)}}}{2}[M_{-1\mid x_{ij^{(i)}}}\rho_{i(j^{(i)}-1)}(M_{-1\mid x_{ij^{(i)}}})^\dagger],\nonumber
\end{eqnarray}

where $\rho_{i0}=\rho^{b_i}_{A_i}$, $F_{ij^{(i)}}$, and $G_{ij^{(i)}}$ are weak measurement parameters and $F_{ij^{(i)}}$ is called the quality factor, which is the undisturbed proportion of the state after $\mathrm{Alice}_{ij^{(i)}}$ is measured, and $G_{ij^{(i)}}$ is called precision factor, which quantifies the information gain through a measurement.

If $j^{(i)}=m$, then $\mathrm{Alice}_{im}$ wants to achieve the maximal correlation with Bob and will perform projective measurements. As a result, the state will change to

\begin{eqnarray}\label{E9}
\rho_{im}=M_{a_{im}\mid x_{im}}\rho_{i(m-1)}(M_{a_{im}\mid x_{im}})^\dagger.
\end{eqnarray}

From the unnormalized postmeasurement state $\rho_{ij^{(i)}}$, we can obtain the probability distribution

\begin{small}
\begin{eqnarray}\label{E10}
P(a_{i1},\dots,a_{ij^{(i)}},b_i\mid x_{i1},\dots, x_{ij^{(i)}},y_i)=\mathrm{Tr}[\rho_{ij^{(i)}}].
\end{eqnarray}
\end{small}

Assume $\mathrm{Alice}_{i1}, i\in\{1,\dots,n\}$ shares a singlet state with Bob, and all Alices and Bob perform the measurements required in the above section. We can derive the quantum upper bound for the network inequality with a combination of arbitrary observers (taking one Alice in each branch)  in an ($n,m,k$) scenario (see Appendix \ref{A} for a proof) as follows 

\begin{eqnarray}\label{E11}
S^{(n,m,k)}_{j}=C_k\left(\prod\limits_{i=1}^n\prod\limits_{o=1}^{j^{(i)}}T_{io}\right)^{1/n},
\end{eqnarray}

where

\begin{eqnarray}
T_{io}=\left\{	\begin{aligned}	
	1 \qquad o=j^{(i)}=m\nonumber\\
	G_{io} \quad o=j^{(i)}<m\nonumber\\
	\frac{1+F_{io}}{2} \qquad\quad o< j^{(i)}\nonumber\\
\end{aligned}
\right.
\end{eqnarray}

This bound only depends on the number of inputs $k$ and the weak measurement parameters of the involved Alices (except the last Alice in every branch) in a star network. In the residual of this paper, we mark this bound as $S_j$ when $(n,m,k)$ is given.

\begin{figure*}
	\centering
	\includegraphics[width=6.5in]{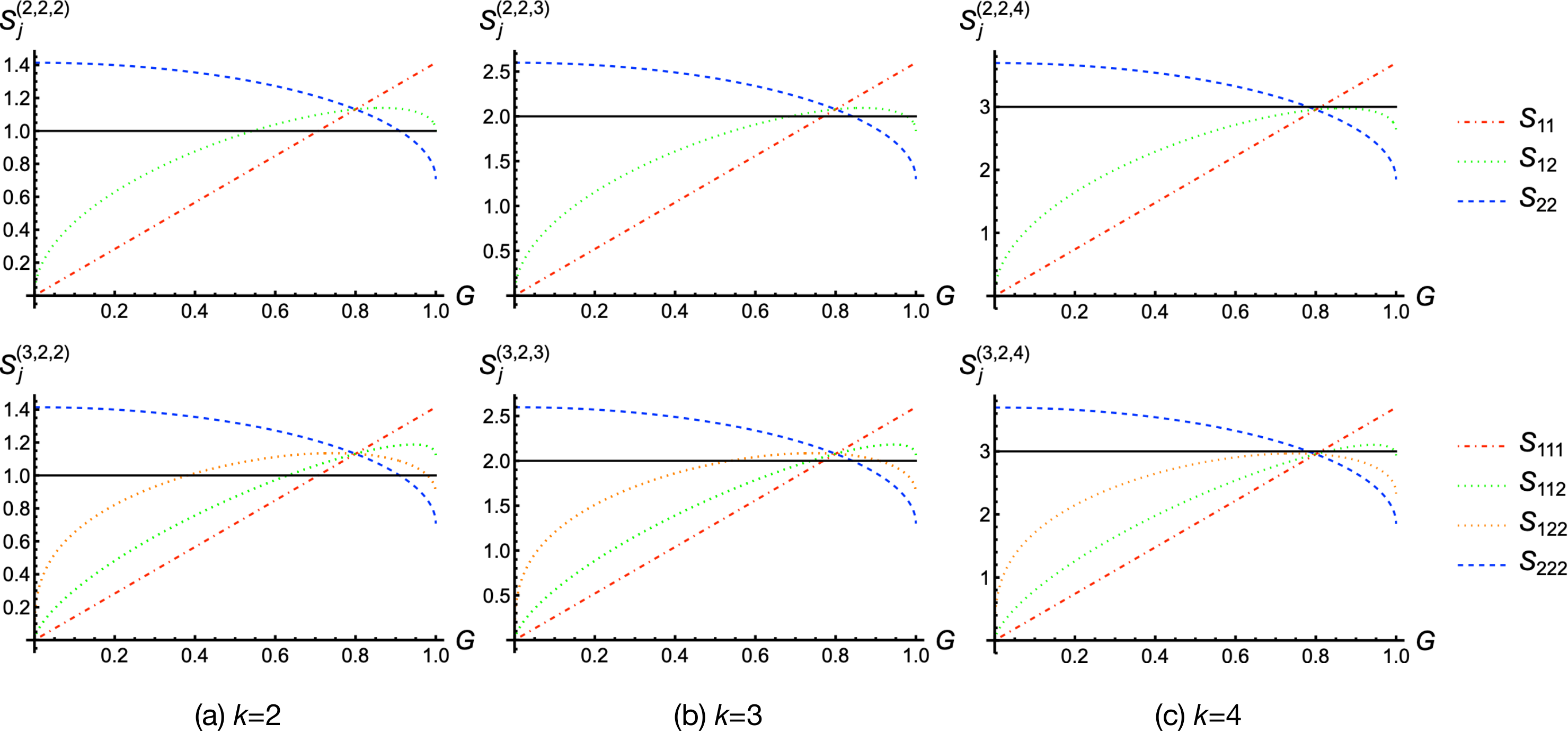}
	\vspace{0.04cm}
	\centering
	\caption{\small{Plot of $S_{j}\in\{S_{11}, S_{12}, S_{21}, S_{22}\}$  in the $(2,2,k)$ scenarios (upper subfigure, $k\in\{2,3,4\}$) with red dot-dashed line, green dotted line ($S_{12}=S_{21}$) and blue dashed line, respectively, and $S_{j}\in\{S_{111}, S_{222}, S_{112}, S_{121}, S_{211}, S_{122},  S_{212}, S_{221}\}$ in the $(3,2,k)$ scenario (lower subfigure, $k\in\{2,3,4\}$) denoted by the red dot-dashed line, blue dashed line, green dotted line ($S_{112}=S_{121}=S_{212}$) and orange dotted line ($S_{122}=S_{212}=S_{221}$), respectively, as functions of the precision factor $G$ under the condition of $G_{i1}=G$ , $i\in\{1,\dots,n\}$ for (a) $k$=2; (b) $k$=3; (c) $k$=4.}}
	\label{F2}
\end{figure*}

Let us discuss some special cases of Eq. \eqref{E11}.

(1) $m=1$, i.e., only one Alice in each branch corresponds to the scenario without nonlocality sharing, which has been discussed in Ref. \cite{2018Lee} and $S_{1\dots1}=C_k=k\mathrm{cos}(\frac{\pi}{2k})$. Meanwhile, if we take $k=2$ that backs to the original star network $(n,1,2)$ scenario \cite{2014Tavakoli} and $S_{1\dots1}=\sqrt{2}$. 

(2) In a $(2,2,k)$ scenario, $S_{j^{(1)}j^{(2)}}$ corresponding to $\mathrm{Alice}_{1j^{(1)}}-\mathrm{Alice}_{2j^{(2)}}-\mathrm{Bob}$ is calculated as follows:

\begin{eqnarray}\label{E12}
	&&S_{11}=C_k\sqrt{G_{11}G_{21}},\ S_{12}=\frac{C_k}{\sqrt{2}}\sqrt{G_{11}(1+F_{21})},\\
	&&S_{21}=\frac{C_k}{\sqrt{2}}\sqrt{G_{21}(1+F_{11})},\ S_{22}=\frac{C_k}{2}\sqrt{(1+F_{11})(1+F_{21})}.\nonumber
\end{eqnarray}

Note that when $k=2$, Eq. \eqref{E12} is coincident with the result in \cite{2022Hou}.

(3) In a generalized $(n,m,k)$ scenario, $S_{1\dots1}=C_k(G_{11}\dots G_{n1})^{1/n}$ and $S_{m\dots m}=C_k(\frac{1+F_{11}}{2}\dots\frac{1+F_{n1}}{2}\dots\frac{1+F_{1(m-1)}}{2}\dots\frac{1+F_{n(m-1)}}{2})^{1/n}$

For simplicity, we assume that $G_{ij^{(i)}}$ and $F_{ij^{(i)}}$ are symmetric in the following section, i.e., $G_{it}=G_t$ and $F_{it}=F_t, i\in\{1,\dots,n\},t\in\{1,\dots,m-1\}$. Then, $S_{1\dots1}$ and $S_{m\dots m}$ in case (3) are reduced to $C_kG_1$ and $C_k(\frac{1+F_{1}}{2}\dots\frac{1+F_{(m-1)}}{2})$, respectively.

\section{Cases of the optimal trade-off relation of weak measurement parameters for different numbers of input}\label{S4}

According to the relationship between the quality factor $F$ and precision factor $G$, there is an optimal trade-off between $F$ and $G$ \cite{2015Silva}, i.e., $F^2+G^2=1$, where $G,F\in[0,1]$, and optimal means by a measurement from which the most information can be extracted with the same disturbance.

In a generalized $(n,2,k)$ scenario, we take $G_1=G, F_1=F$. The quantum upper bound is given as $S_j=C_k[G^{n_1}(\frac{1+F}{2})^{n_2}]^{1/n}$, where $n_1$ ($n_2$) means the number of 1 (2) in $j$ and $n_1+n_2=n$. In total, $2^n$ quantities can be discussed; in particular, $S_{1\dots1}= C_kG$ and $S_{2\dots 2}=C_k
\frac{1+F}{2}$. 

We analyze the optimal relation of weak measurement parameters in the following. Obviously, the violation intervals of all $2^n$ inequalities with respect to $G$ are only determined by $S_{1\dots1}$ and $S_{2\dots 2}$ together. When $k=2$, these $2^n$ quantities acquired by Eq. \eqref{E11} can simultaneously exceed the classical bound 1 in a range of $G\in\{\frac{1}{\sqrt{2}},\sqrt{2(\sqrt{2}-1)}\}$, and the maximum simultaneous violation value is $\frac{4\sqrt{2}}{5}\approx1.13137$ when $G=0.8$. When $k$=3, the quantum upper bounds acquired by Eq. \eqref{E11} can simultaneously exceed the classical bound 2 in a relatively narrow range of $G\in\{\frac{4}{3\sqrt{3}},\frac{4}{3}\sqrt{\frac{1}{3}(3\sqrt{3}-4)}\}$, and the maximum simultaneous violation value is $\frac{6\sqrt{3}}{5}\approx2.07846$ when $G=0.8$. No simultaneous violation of all quantities in a $(n,2,k)$ scenario occurs when $k>3$, and when $k=4, G=0.8$, these quantities acquired by Eq. \eqref{E11} simultaneously achieve a maximum value of $\frac{8}{5}\sqrt{2+\sqrt{2}}\approx2.9564$, which is lower than the classical bound of 3. 

\begin{figure*}[!htbp]
	\includegraphics[width=6in]{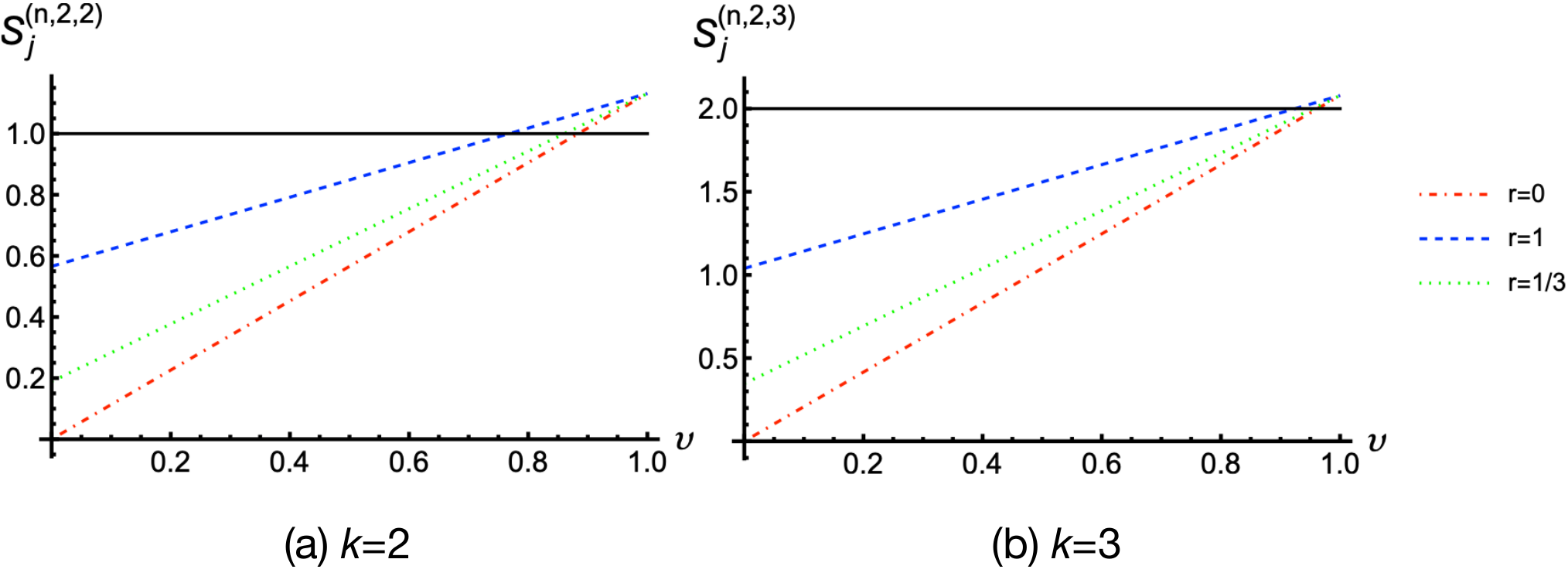}
	\vspace{0.04cm}
	\centering
	\caption{\small{Comparison of the different noise types affecting $S^{(n,2,k)}_{j}$ when $G$=0.8. Plot of $S_{j}$ with the visibility $v$ for (a) $k=2$ and (b) $k=3$, where the red dot-dashed line, blue dashed line and green dotted line represent the scenarios of $r=0$ (white noise), $r=1$ (colored noise) and $r=1/3$, respectively. The violation is more sensitive to white noise. }}
	\label{F3}
\end{figure*}

These quantum upper bounds in the $(2,2,k)$ and $(3,2,k)$ scenarios ($k\in\{2,3,4\}$) can be depicted as a function of $G$ shown in Fig. \ref{F2}. In the $(2,2,4)$ scenario, as depicted in the upper subfigure of Fig. \ref{F2}(c), when $G=\frac{\sqrt{3}}{2}\approx0.8660$, $S_{12}$ and $S_{21}$ achieve the maximal value $3^{3/4}\sqrt{\frac{2 +\sqrt{2}}{2}}\approx2.9783$, which is lower than the classical bound of 3. In the $(3,2,4)$ scenario as depicted in the lower subfigure of Fig. \ref{F2}(c), when $G=\frac{\sqrt{5}}{3}\approx0.7453$, the quantities $S_{122},S_{212},S_{221}$ achieve the maximal value $\frac{2^{1/3}5^{5/6}\sqrt{2+\sqrt{2}}}{3}\approx2.9671$, which is lower than the classical bound of 3. Simultaneous violation only exists among $S_{111},S_{112},S_{121},S_{211}$ in a range of $G\in\{0.8280,0.9970\}$, and the maximal simultaneous violation $\frac{2^{10/3}}{6}\sqrt{2+\sqrt{2}}\approx3.1040$ is achieved when $G=\frac{2\sqrt{2}}{3}\approx0.9428$. 

The conclution that there is no nonlocality sharing among all observers in the generalized star network when $m=3$ can be immediately derived. In this $(n,3,k)$ scenario, $S_{1\dots1}=C_kG_1$, $S_{2\dots2}=C_k\frac{1+F_1}{2}G_2$, and $S_{3\dots3}=C_k\frac{1+F_1}{2}\frac{1+F_2}{2}$. These three quantities achieve the maximal value at the same time as $C_k\cdot\frac{20}{29}$ when $G_1=\frac{20}{29}$ and $G_2=0.8$; when $k=2$, this value is $\frac{20\sqrt{2}}{29}\approx0.9753$, which is lower than the classical bound of 1.

\section{Noise resistance in network nonlocality sharing}\label{S5}

From an experimental perspective, producing perfect maximally entangled states is an extremely demanding task. Therefore, it is natural to consider that imperfect particles exist in the network and discuss the influence on the network correlations. In the most common photonic experiments, entangled photons are usually produced through a spontaneous parametric down conversion (SPDC) process in a nonlinear crystal, and two different classes of noise affect the SPDC sources: white noise and colored noise \cite{2005Cabello,2020FabioS}. Suppose these sources distribute the singlet state as previously discussed, and the noise states can be expressed as

\begin{equation}\label{E13}
	\rho_i^{\mathrm{w}}=v_i\ket{\psi^-}\bra{\psi^-}+(1-v_i)\frac{\mathbb{I}}{4},
\end{equation}

and
 
\begin{equation}\label{E14}
	\rho_i^{\mathrm{c}}=v_i\ket{\psi^-}\bra{\psi^-}+(1-v_i)M_{\mathrm{color}},
\end{equation}

where $M_{\mathrm{color}}$ =$\frac{1}{2}(\ket{01}\bra{01}+\ket{10}\bra{10})$ describes the depolarization direction and the colored noise is intrinsic in the SPDC process.

Consequently, the final state can be modeled by combining these two different contributions in a normalized form:

\begin{eqnarray}\label{E15}
	\rho_i&&=(1-r_i)\rho_i^{\mathrm{w}}+r_i\rho_i^{\mathrm{c}},
\end{eqnarray}

where $v_i$ and $r_i$ represent the total noise and the fraction of colored noise of source $\rho_i$, respectively.

We focus on the quantum upper bounds among $\mathrm{Alice}_{11}-\dots-\mathrm{Alice}_{n1}$-Bob and $\mathrm{Alice}_{12}-\dots-\mathrm{Alice}_{n2}$-Bob in the cases of where $k=2$ and $k=3$.

If $r_i=0$, which corresponds to there being only white noise, we obtain

\begin{eqnarray}\label{E16}
	&&S^{(n,2,k)}_{1\dots1}=C_k\left\{\prod\limits_{i=1}^n G_{i1}v_i\right\}^{1/n},\\
	&&S^{(n,2,k)}_{2\dots2}= \frac{C_k}{2}\left\{\prod\limits_{i=1}^n (1 + F_{i1}) v_i\right\}^{1/n}.\nonumber
\end{eqnarray}

V=$(\prod\limits_{i=1}^n v_i)^{1/n}$ is the critical visibility for network nonlocality sharing in this situation. When $k=2$ and $k=3$,  there is no simultaneous violation of network inequality between $\mathrm{Alice}_{11}-\dots-\mathrm{Alice}_{n1}$-Bob and $\mathrm{Alice}_{12}-\dots-\mathrm{Alice}_{n2}$-Bob with $V\leq88.39\%$ and $V\leq96.23\%$, respectively.

In the following, we consider that the noise parameters and the weak measurement parameters are the same for all of the sources and observers $\mathrm{Alice}_{i1}$, respectively, i.e., $r_1=\dots=r_n=r$, $v_1=\dots=v_n=v$ and $G_{11}=\dots G_{n1}=G$, respectively. The weak measurement parameter relations of the corresponding Alices are all optimal; then, these quantum upper bounds are reduced to

\begin{eqnarray}\label{E17}
	&&S^{(n,2,k)}_{1\dots1}=\frac{C_k}{2}\left\{ G \left[r (1-v) + 2v\right]\right\},\\
	&&S^{(n,2,k)}_{2\dots2}= \frac{C_k}{4}\left\{(1 + \sqrt{1-G^2}) [r (1-v) + 2v]\right\}.\nonumber
\end{eqnarray}

Therefore, we can analyze the influence of different noise types on the network correlation. As illustrated in Fig. \ref{F3}, we list the quantities in Eq. \eqref{E17} for different number of inputs when $G$=0.8, i.e., the parameter achieves the maximal violation. We find that the violation is more sensitive to white noise. The concrete proportion of parameter $r$ can be evaluated in the experiment \cite{2020FabioS}. For the case when $r=1/3$, the critical visibility for $k=2$ and $k=3$ is $86.07\%$ and $95.48\%$, respectively.

The generalized expression for when all of the sources contain different noise parameters is provided in Appendix \ref{B}.

\section{Summary and discussion}\label{S6}

In this paper, the effect of different numbers of inputs on network nonlocality sharing in a generalized star network configuration is discussed. In the cases where $m=2$ and the numbers of input are $k=2$ or $k=3$, all $2^n$ star network inequalities with respect to $\mathrm{Alice}_{1j^{(1)}}-\dots-\mathrm{Alice}_{nj^{(n)}}-\mathrm{Bob}$ can be simultaneously violated for any $n$ branches. A natural extension to our study would be to reveal network nonlocality sharing among all observers when $k>3$, and we propose two probable ways to achieve this proposal. The first possible scheme is to utilize other multisetting star network inequalities such as the Munshi-Kumar-Pan (MKP) inequality proposed in Ref. \cite{2021Munshi}. In each branch, the former Alice must perform weaker measurements to decrease her violation for the latter Alice to achieve nonlocal correlation with Bob. As the number of inputs $k$ increases, the ratio of the quantum upper bound and classical bound of the MKP inequality approaches to 1.25 when high-dimensional systems are used. A second possible direction is using the method of unequal sharpness measurements. Considering different sharpness parameters for two different measurements by each observer can significantly increase the number of observer who can achieve nonlocality sharing \cite{2020BrownP}, and it is interesting to study whether this phenomenon still holds for a larger number of input.

From an experimental perspective, the separable measurement is the simplest possible measurement in practice. Our result for the star network nonlocality sharing is experimentally observable, and there are some reference experiments: star network nonlocality has been observed in the recent experiments \cite{2020FabioS,2021ZhangC} and the weak measurement part used in this paper has been implemented in
several experiments \cite{2018Hu,2017Matteo,2020FengTF,2020Foletto}.

\begin{acknowledgments}
	J.H.W acknowledges the useful discussions with Qingsong Chang. We thank A. K. Pan for helpful comments. This work is supported by the National Natural Science Foundation of China (Grant Nos. 12165020, 11575155) and the Major Science and Technology Project of Yunnan Province, China (Grant No. 2018ZI002).
\end{acknowledgments}

\appendix

\section{Proof of Eq. (11)}\label{A}

According to the expression of Eq. \eqref{E5}, we first need to calculate the probability distribution $P(a_{i1},\dots,a_{ij^{(i)}},b_i\mid x_{i1},\dots, x_{ij^{(i)}},y_i)$.

If $j^{(i)}=m$, then according to Eq. \eqref{E9}, we have

\begin{eqnarray}
&&\sum\limits_{a_{im}}a_{im}\mathrm{Tr}[\rho_{im}]\\
&&=\mathrm{Tr}\left\{[\mathrm{sin}(\frac{x_{im}\cdot\pi}{k})\sigma_x+\mathrm{cos}(\frac{x_{im}\cdot\pi}{k})\sigma_z]\rho_{i(m-1)}\right\}.\nonumber
\end{eqnarray}

If $j^{(i)}<m$, then according to Eq. \eqref{E8}, we have

\begin{small}
\begin{eqnarray}
&&\sum\limits_{a_{ij^{(i)}}}a_{ij^{(i)}}\mathrm{Tr}[\rho_{ij^{(i)}}]=G_{ij^{(i)}}\\
&&\times \mathrm{Tr}\left\{[\mathrm{sin}(\frac{x_{ij^{(i)}}\cdot\pi}{k})\sigma_x+\mathrm{cos}(\frac{x_{ij^{(i)}}\cdot\pi}{k})\sigma_z]\rho_{i(j^{(i)}-1)}\right\}.\nonumber
\end{eqnarray}
\end{small}

Recursively compute these states until $\rho_{i0}=\rho^{b_i}_{A_i}$; for example,

\begin{scriptsize}
\begin{eqnarray}
&&\frac{1}{k}\sum\limits_{\mbox{\tiny $\begin{subarray}{l}a_{i(j^{(i)}-1)}\\x_{i(j^{(i)}-1)}\end{subarray}$}}\mathrm{Tr}\left\{[\mathrm{sin}(\frac{x_{ij^{(i)}}\cdot\pi}{k})\sigma_x+\mathrm{cos}(\frac{x_{ij^{(i)}}\cdot\pi}{k})\sigma_z]\rho_{i(j^{(i)}-1)}\right\}\\
&&=\frac{1+F_{i(j^{(i)}-1)}}{2}\mathrm{Tr}\{[\mathrm{sin}(\frac{x_{ij^{(i)}}\cdot\pi}{k})\sigma_x+\mathrm{cos}(\frac{x_{ij^{(i)}}\cdot\pi}{k})\sigma_z]\rho_{i(j^{(i)}-2)}\}.\nonumber
\end{eqnarray}
\end{scriptsize}

When $\rho_i$ is a singlet state, we have

\begin{eqnarray}
&&\sum\limits_{b_i}b_i\mathrm{Tr}\left\{[\mathrm{sin}(\frac{x_{ij^{(i)}}\cdot\pi}{k})\sigma_x+\mathrm{cos}(\frac{x_{ij^{(i)}}\cdot\pi}{k})\sigma_z]\rho^{b_i}_{A_i}\right\}\nonumber\\
&&=-[\mathrm{cos}(\frac{\pi(1-2x_{ij^{(i)}}+2y)}{2k})].
\end{eqnarray}

Consequently, the correlator is calculated as

\begin{small}
\begin{eqnarray}
\langle A_{ij^{(i)}}^{x_{ij^{(i)}}}B^{i-1}_i\rangle&&=-\prod\limits_{o=1}^{j^{(i)}}T_{io}\cdot \mathrm{cos}(\frac{\pi(1-2x_{ij^{(i)}}+2(i-1))}{2k}),\nonumber\\
\\
&&T_{io}=\left\{	\begin{aligned}	
	1  \qquad o=j^{(i)}=m\\
	G_{io} \quad o=j^{(i)}<m\\
	\frac{1+F_{io}}{2} \qquad\quad o< j^{(i)}\nonumber\\
\end{aligned}
\right.
\end{eqnarray}
\end{small}

Therefore, in a generalized star network $(n,m,k)$ scenario with the $n$-branch, $m$ observers in each branch and $k$ inputs, the quantum upper bound (shorthand $A_{ij^{(i)}}^{x_{ij^{(i)}}}$ as $A^i$) is given by

\begin{small}
\begin{eqnarray}
S&&^{(n,m,k)}_{j}=\sum\limits_{l=1}^k\vert \mathcal{I}_{lj}\vert^{1/n},\nonumber\\
&&=\frac{1}{2}\sum\limits_{l=1}^k\left(\vert\sum\limits^l_{x_{1j^{(1)}},\dots,x_{nj^{(n)}}=l-1}\langle A^1,\dots,A^nB^{l-1}\rangle\vert\right)^{1/n},\nonumber\\
&&=\frac{1}{2}\sum\limits_{l=1}^k\left(\vert\sum\limits^l_{x_{1j^{(1)}}=l-1}\langle A^1B_1^{l-1}\rangle\dots \sum\limits^l_{x_{nj^{(n)}}=l-1}\langle A^nB^{l-1}_n\rangle\vert\right)^{1/n},\nonumber\\
&&=k\mathrm{cos}(\frac{\pi}{2k})\left(\prod\limits_{i=1}^n\prod\limits_{o=1}^{j^{(i)}}T_{io}\right)^{1/n}.
\end{eqnarray}
\end{small}

\section{Quantum upper bound for noise states}\label{B}

When $\rho_i$ is the state in Eq. \eqref{E15}, we have 

\begin{small}
\begin{eqnarray}
	&&\sum\limits_{b^i}b^i\mathrm{Tr}\left\{[\mathrm{sin}(\frac{x_{ij^{(i)}}\cdot\pi}{k})\sigma_x+\mathrm{cos}(\frac{x_{ij^{(i)}}\cdot\pi}{k})\sigma_z]\rho^{b_i}_{A_i}\right\}=\\
	&&-[v_i\mathrm{cos}(\frac{\pi(1-2x_{ij^{(i)}}+2y)}{2k})+(1-v_i)r_i\mathrm{cos}\frac{x_{ij^{(i)}\pi}}{k}\mathrm{cos}\frac{(2y+1)\pi}{2k}].\nonumber
\end{eqnarray}
\end{small}

Consequently, the correlator is calculated as

\begin{small}
	\begin{eqnarray}
		\langle A_{ij^{(i)}}^{x_{ij^{(i)}}}B^{i-1}_i\rangle&&=-\prod\limits_{o=1}^{j^{(i)}}T_{io}\cdot [v_i\mathrm{cos}(\frac{\pi(1-2x_{ij^{(i)}}+2y)}{2k})+\nonumber\\
		&&(1-v_i)r_i\mathrm{cos}\frac{x_{ij^{(i)}}\pi}{k}\mathrm{cos}\frac{(2y+1)\pi}{2k}].
	\end{eqnarray}
\end{small}

After a direct calculation, we have

\begin{small}
\begin{eqnarray}
	&&S^{(n,2,k)}_{1\dots1}=\frac{C_k}{2}\left\{\prod\limits_{i=1}^n G_{i1} \left[r_i (1-v_i) + 2v_i\right]\right\}^{1/n},\\
	&&S^{(n,2,k)}_{2\dots2}= \frac{C_k}{4}\left\{\prod\limits_{i=1}^n (1 + F_{i1}) [r_i (1-v_i) + 2v_i]\right\}^{1/n}.\nonumber
\end{eqnarray}
\end{small}

\end{document}